\documentclass[aps,prl,showpacs,twocolumn,superscriptaddress,longbibliography,amsmath,amssymb]{revtex4-1}
\usepackage[pdftex]{graphicx}
\usepackage[pdftex,bookmarks,colorlinks,breaklinks]{hyperref} % PDF hyperlinks, with coloured links
\usepackage{amsmath,amssymb}
\usepackage{bbold}
\usepackage{color}
\hypersetup{linkcolor=red,citecolor=blue,filecolor=blue,urlcolor=blue} % coloured links
\hypersetup{pdftoolbar=true,pdfpagemode=UseNone,pdfstartview=FitH,colorlinks=true,extension=}

\newcommand{\CT}{\ensuremath{{\cal{CT}}}}
\newcommand{\T}{\ensuremath{{\cal{T}}}}
\newcommand{\C}{\ensuremath{{\cal{C}}}}
\newcommand{\G}{\ensuremath{{\cal{G}}}}
\newcommand{\Hcal}{\ensuremath{{\mathcal{H}}}}

\newcommand{\ket}[1]{\ensuremath{{\left| #1 \right\rangle}}}
\newcommand{\bra}[1]{\ensuremath{{\left\langle #1 \right|}}}
\newcommand{\braket}[2]{\ensuremath{{\left\langle\left. #1 \right| #2 \right\rangle}}}
\renewcommand{\Re}[1]{\ensuremath{ \textrm{Re}\left(#1\right) }}

\newcommand{\diff}[2]{\frac{d#1}{d#2}}
\newcommand{\e}{\textrm{e}}

\begin{document}
\title{Non-Hermitian \CT-Symmetric Spectral Protection of Nonlinear Defect Modes}
\author{Do Hyeok Jeon}
\affiliation{Wave Transport in Complex Systems Lab, Physics Department, Wesleyan University, Middletown CT-06459, USA}
\author{Mattis Reisner}
\affiliation{Universit\'{e} C\^{o}te d'Azur, CNRS, Institut de Physique de Nice (INPHYNI), 06108 Nice, France, EU}
\author{Fabrice Mortessagne}
\affiliation{Universit\'{e} C\^{o}te d'Azur, CNRS, Institut de Physique de Nice (INPHYNI), 06108 Nice, France, EU}
\author{Tsampikos Kottos}
\affiliation{Wave Transport in Complex Systems Lab, Physics Department, Wesleyan University, Middletown CT-06459, USA}
\author{Ulrich Kuhl}
\affiliation{Universit\'{e} C\^{o}te d'Azur, CNRS, Institut de Physique de Nice (INPHYNI), 06108 Nice, France, EU}

\begin{abstract}
We investigate, using a microwave platform consisting of a non-Hermitian Su-Schrieffer-Heeger array of coupled dielectric resonators, the interplay of a lossy nonlinearity and \CT-symmetry in the formation of defect modes. 
The measurements agree with the theory which predicts that, up to moderate pumping, the defect mode is an eigenstate of the \CT-symmetric operator and retains its frequency at the center of the gap. 
At higher pumping values, the system undergoes a self-induced explicit \CT-symmetry violation which removes the spectral topological protection and alters the shape of the defect mode. 
\end{abstract}

\maketitle

\emph{Introduction -- } 
Topological photonics (TP) \cite{lu14,oza19} has originally developed within the framework of Hermitian wave physics, drawing inspirations from the discovery of exotic topological phases appearing in traditional condensed matter. 
Its rapid blossom relies on the promise that the developed TP methodologies, based on geometrical and topological concepts, can lead to unprecedented control on light-matter interactions. 
A prominent application is the realization of photonic structures with transport characteristics that are immune to fabrication imperfections \cite{hal08,wan09,rec13,haf13}. 
Most of these investigations principally focus on linear topological phenomena. 
Recently, however, it has been realized that nonlinearities, when invoked, can offer a dynamical tuning mechanism that produces a variety of exotic phenomena. 
Examples include robust discrete solitons \cite{had17}, self-localized topological edge solitons \cite{lum13,ley16}, topologically enhanced harmonic generation \cite{wan19}, optical isolation \cite{zhu17}, topological lasers \cite{mal18a}, and self-induced topological states \cite{had16,had18,dob18}. 

At the same time, the topological physics agenda (set initially by the condensed matter community) has been enlarged and redefined by the necessities presented in the optics framework. 
In particular, the natural presence of non-Hermitian elements like gain and loss ``demands'' the re-definition or (even!) invention of new topological concepts and classifications. 
It was, therefore, only natural that the interplay between non-Hermiticity and topological protection attracted recently a lot of theoretical and experimental interest  \cite{sch13,ZRPLNRSS15,pol15,WKPLNMSRS16,L16,LBHCN17,kuh17,mak18,SZF18,MBBF18,HBLRCKCS18,BWHPRSCK18,ZMTMGSF18,GAKTHU18,BS18,kaw19,rei19,arXrei19a}. 
Some of the questions that have been raised include the generalization of band-edge correspondence, the emergence of new topological states without any Hermitian counterpart, or the necessity of a new topological phase classification \cite{oza19,MBBF18,GAKTHU18,kaw19}. 
On the technological side, the development of novel classes of topologically protected lasers \cite{HBLRCKCS18,BWHPRSCK18,ZMTMGSF18} and reflective photonic limiters \cite{kuh17,rei19,arXrei19a} introduced a new excitement and the urge to understand better the formation and spectral stability of topological states in the framework of non-Hermitian physics.

Here we analyze, both experimentally and theoretically, the intricate effects that nonlinearity together with non-Hermiticity have in the formation of a topological defect mode. 
Our platform utilizes a standard Su-Schrieffer-Heeger (SSH) binary array consisting of identical microwave resonators coupled electromagnetically with one another. 
In the middle of the array we position one defect resonator which involves both linear and nonlinear losses, thus enforcing a charge-conjugation (\CT) symmetry to the whole SSH structure. 
We find that the emerging defect mode is an eigenmode of the \CT-operator. 
Furthermore, under pump-probe measurements, this mode is resilient to a large range of pumping powers, with its frequency being pinned at the center of the band-gap. 
For even higher values of the pump, the system experiences a self-induced explicit \CT-symmetry violation which enforces the destruction of the defect mode. 
The measurements are in excellent agreement with the results from a theoretical analysis which utilizes a modified non-Hermitian Green's function's formalism that treats the non-Hermitian nonlinear defect perturbatively. 
Our results paves the way for the design of topologically protected isolators, circulators, or switches with self-induced reconfigurability.

\begin{figure}
	\centering
	\includegraphics[width=.99\columnwidth]{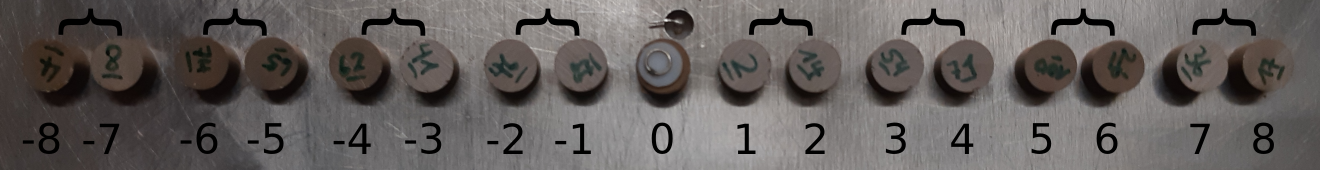}
	\caption{\label{fig:fig1_photo_set_up}
		The experimental setup. 
		The dielectric resonators are placed on an alumina plate and form an SSH binary structure, where the dimers have a small distance $d_1=10$\,mm corresponding to the strong coupling $t_1=68$\,MHz (indicated by the braces). 
		The distance between dimers is $d_2=12$\,mm (weak coupling $t_2=33$\,MHz). 
		The central resonator ($m=0$) is weakly coupled to both of its neighboring resonators and on top of it the short circuited diode is seen on a teflon spacer.
		The kink antenna as the excitation antenna is positioned at the defect resonator. 
		The whole system is closed by a top plate which supports also the scanning loop antenna (not shown).
	}
\end{figure}

\emph{SSH CROW array and its mathematical modeling -- } 
The simplest platform on which we can examine the interplay of nonlinearities with non-Hermiticity and topological protection is the SSH model \cite{su79}. 
It consists of a one-dimensional (1D) periodic array of identical resonators with alternating short ($d_1$) and long ($d_2$) distances from one another. 
A defect resonator is introduced in the middle of the chain, by placing two consequent resonators in long distance from one another. 
Furthermore we assume that the defect resonator supports linear and nonlinear losses. 
We have implemented experimentally this set-up by realizing a transmission line consisting of coupled resonator microwave waveguide (CRMW) array, see Fig.~\ref{fig:fig1_photo_set_up}. 
The array consists of $N=17$ high-index ($n_r=6$) cylindrical resonators (radius $r=4$\,mm, height $h=5$\,mm) made of ceramics (ZrSnTiO) with resonant frequency around $\epsilon\approx 6.876$\,GHz.
The distances between the resonators is $d_1=10$\,mm and $d_2=12$\,mm corresponding to strong $t_1=68$\,MHz and weak $t_2=33$\,MHz electromagnetic coupling respectively. 
Since nonlinear material properties in the microwave domain are very weak, we have incorporated nonlinearity by coupling inductively the central resonator with a diode (Detector Schottky Diode SMS7630-079LF from Skyworks). 
The latter has been placed above the defect resonator using a spacer of Teflon. 
The diode is short circuited and coupled via a metallic ring with a diameter of $3$\,mm. 
Thus the $z$-directional magnetic field of the electromagnetic radiation at the defect resonator is inductively coupled to the fast diode. 
Finally, the system is pumped externally via a vector network analyzer (ZVA~24 from Rohde \& Schwarz) which is set to inject a power of $P_\textrm{VNA}$ ranging from -200 to 10\,dBm via the strongly coupled kink antenna (see Fig.~\ref{fig:fig1_photo_set_up}). 
Note that this is not the real applied power on the defect resonator as there are additional absorption in the cables, the coupling of the antenna to the defect resonator, and other loss types, which are all of linear type. 
Instead, the field intensity at the defect resonator $I_{D}$ is proportional to the injected power, $I_{D}=a_\textrm{lin} P_\textrm{VNA}$ in the steady state situation realized by the experiment. 
The field is probed by a weakly coupled loop antenna fixed on the top plate which is movable. 
For details on the different kind of antennas, experimental setup see \cite{rei19}.

\begin{figure}
\centering
\includegraphics[width=.99\columnwidth]{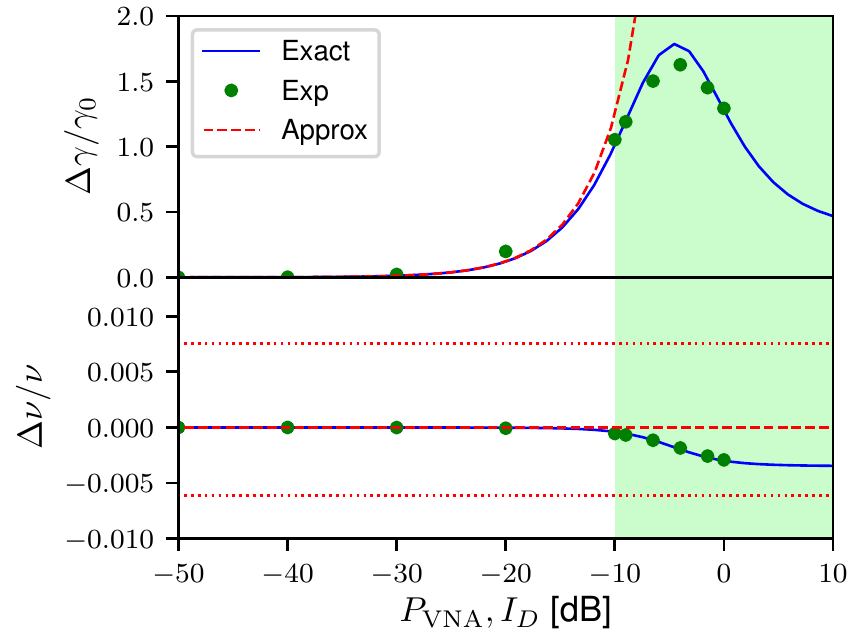}
\caption{\label{fig:fig2_omega_shift}
Spectral shift of the complex defect mode frequency $\omega_D = \nu_D + i \gamma_D$ with varying injected power $P_\textrm{VNA}$ (expressed as dBm) at the defect site. 
Note that shifts in the real and imaginary parts of the eigenfrequency have each been renormalized with respect to its low $P_\textrm{VNA}$ limit, $\nu_0$ and $\gamma_0$, respectively. 
Measurements (green dots) of the eigenfrequency show agreement with the theoretical results (blue curves), with $\Omega(I_D)$ given by the first expression in Eq.~(\ref{eq:Omega_sat}). 
From low to moderate $P_\textrm{VNA}$, the saturable nonlinearity can be well approximated by the second expression in Eq.~(\ref{eq:Omega_sat}), from which we have also calculated the complex eigenfrequency (red dashed curves). 
The red dotted lines signify the upper and lower limits of the band gap, extracted from the measured transmission spectra.
}
\end{figure}

The above experimental set-up is described mathematically using the following couple mode theory (CMT):
\begin{eqnarray}\label{eq:CMTeqs} 
\nonumber
\Hcal \ket{\Psi_l}&=&\omega_l \ket{\Psi_l}, \qquad \ket{\Psi_l}=\sum_m\psi_m^{(l)}\ket{m}, \\ 
\nonumber
\Hcal &=& \sum_m \ket{m}\epsilon_m \bra{m} + \ket{m}t_{m,m-1} \bra{m-1} \\
&& \hspace{6em} + \ket{m} t_{m,m+1} \bra{m+1},
\end{eqnarray}
where $\Hcal$ is the effective Hamiltonian that describes the system, $\omega_l$ is the $l$-th eigenfrequency of the SSH and $\psi_m^{(l)}$ represents the corresponding magnetic field amplitude of the $l$-th super-mode in the individual resonator (Wannier) basis $\ket{m}$ localized at the $m=[-(N-1)/2],\cdots,[(N-1)/2]$-th resonator (site). 
The resonant frequency of the $m$-th resonator is indicated as $\epsilon_m=\epsilon$ (for $m\neq 0$) and $t_{m,m+1}$ is the coupling coefficient between $m$-th and $(m+1)$-th resonators. 
The defect resonator is at position $m\equiv m_D=0$ at the center of the array (see Fig.~\ref{fig:fig1_photo_set_up}). Hamiltonian~(\ref{eq:CMTeqs}) (in the absense of nonlinearities and $N\rightarrow \infty$) has a band structure with $\omega (k)=\epsilon\pm\sqrt{t_1^2+t_2^2+2t_1t_2\cos(k)}$ ($k\in[-\pi;\pi]$ is the wavenumber) and one defect mode at $\omega_D=\epsilon$ (center of the gap). 
The nonlinearity due to the presence of the PIN diode has been incorporated in Eq.~(\ref{eq:CMTeqs}) by modifying the resonant frequency $\epsilon_D$ as
\begin{equation}\label{eq:eq_nonl}
\epsilon_D = \epsilon + \Omega(I_D), 
\end{equation}
where $\Omega(I_D)$ is a nonlinear function of the local field intensity $I_D=|\psi_D|^2$ at the defect site $m_D$. 

In case where $\Omega(I_D)$ is purely imaginary, the system described above respects a charge-conjugation (\CT) symmetry, defined as anti-commutation of the Hamiltonian $\Hcal$ with the $\CT$-operator. 
Here, $\T$ is the time-reversal operator (equivalent to the complex conjugation) and $\C$ is the chiral symmetric operator, which in the Wannier basis takes the form $\bra{m_1}\C\ket{m_2}=(-1)^{m_1} \delta_{m_1,m_2}$. 
Charge conjugation symmetry imposes special restrictions to the spectrum of the system; namely the eigenstates of a $\CT$ symmetric Hamiltonian $\Hcal$ come in pairs, known as $\CT$ symmetric partners. 
In particular, given an eigenstate $\Hcal\ket{\Psi}=\omega\ket{\Psi}$, we can obtain its \CT partner $\Hcal\CT\ket{\Psi}=-\omega^*\CT\ket{\Psi}$. 
This implies that the spectral features of $\Hcal$ is mirror-symmetric with respect to the resonant frequency of the individual resonators. 
An eigenstate whose \CT partner is distinct from itself is referred to be in {\it broken \CT-symmetric phase}. 
On the other hand, an eigenstate in the {\it exact \CT-symmetric phase} is the $\CT$ symmetric partner of itself, $\CT\ket{\Psi}=\ket{\Psi}$. 
It follows that an exact \CT-symmetric state has purely imaginary eigenfrequency shift i.e. ($\Re{\omega}-\epsilon=0$). 
Moreover, a global complex phase in the wave function of such a state can be chosen such that purely real (imaginary) components occupy the even(odd)-numbered sublattices \cite{note1}.

\emph{Theoretical analysis of defect mode -- } 
To investigate the symmetry-induced topological features of the defect mode we developed a non-Hermitian nonlinear Green's function formalism, which is in general applicable to non-Hermitian systems with nonlinear defects. 
To this end, we decompose the total Hamiltonian $\Hcal$ into an unperturbed Hermitian $\Hcal_0$ and a non-Hermitian perturbation $\Hcal_1$ which describes the lossy nonlinearity at site $m_D$. 
Specifically
\begin{equation}
\Hcal=\Hcal_0+\Hcal_1; \quad \Hcal_1=\ket{D}\Omega(I_D)\bra{D},
\end{equation}
where $\ket{D}$ indicates the Wannier state at the defect site. 
The Green's function $\G(z)\equiv\left(z-\Hcal\right)^{-1}$ of the total Hamiltonian reads \cite{eco79,mol93}
\begin{eqnarray}\label{Gfunction}
\G(z)&=&\G_0\left(1-\Hcal_1\G_0\right)^{-1}\\
&=& \G_0\left(1+\Hcal_1\G_0+\left(\Hcal_1\G_0\right)^2+\cdots\right)\nonumber\\
&=&\G_0+\G_0\T\G_0,\nonumber
\end{eqnarray}
where $\G_0(z)\equiv \left(z-\Hcal_0\right)^{-1}$ is the Green's function of the unperturbed Hamiltonian $\Hcal_0$ and $\T(z)=\ket{D}\frac{\Omega}{1-\bra{D}\G_0\ket{D}\Omega}\bra{D}$ is the so-called $t$-matrix \cite{eco79}. 
From the last line of Eq.~(\ref{Gfunction}) we can evaluate the simple pole $z =\omega_D$ of $\G$ corresponding to the eigenenergy of the defect state. 
The latter is the solution of the following transcendental equation:
\begin{equation}\label{eq:eq1}
\bra{D}\G_0(\omega_D)\ket{D}=\Omega(I_D)^{-1}
\end{equation}
while the corresponding residue is
\begin{equation}\label{eq:eq2}
\textrm{Res}\left[\bra{D}\G(z)\ket{D}\right]_{z=\omega_D}
=\frac{\left(\psi_D^{(D)}\right)^2}{\chi}
=\left[-\frac{\bra{D}\G_0\ket{D}^2}{
	\diff{\bra{D}\G_0\ket{D}}{z}
}\right]_{z=\omega_D}
\end{equation}
where $\psi_D^{(D)}$ is the $m_D=0$-th component of the defect eigenstate (in the Wannier basis). 
As opposed to Hermitian defect perturbation methods \cite{eco79,mol93}, the first equality in Eq.~(\ref{eq:eq2}) requires to use the fact that in non-Hermitian systems the eigenmodes are bi-orthogonal, i.\,e.\ $\sum_l \frac{\ket{\Psi_l}\bra{ \Phi_l}}{\langle \Phi_l|\Psi_l\rangle} = \mathbb{1}$, where $\ket{\Psi_l}$ and $\bra{ \Phi_l}$ denote the $l$-th right and left eigenvectors of \Hcal, respectively. 
In case of non-Hermitian systems, the eigenvectors are normalized as $\langle \Phi_l|\Psi_{l'}\rangle=\sum_m\psi_m^{(l)}\psi_m^{(l')}=\chi \delta_{l,l'}$ where we have used the fact that $\ket{\Psi_l}=\left(\ket{\Phi_l}\right)^*$ in case of symmetric non-Hermitian Hamiltonians $\Hcal=\Hcal^T$. 
The parameter $\chi$ is the so-called quasi-power whose analog in Hermitian physics is the total power of the signal \cite{mak08}.
Equations~(\ref{eq:eq1},~\ref{eq:eq2}) have to be solved self-consistently for $\omega_D$ and $\psi_D^{(D)}$ and for various values of the quasi-power $\chi$ which play the role of a free parameter. 
Using Eq.~(\ref{eq:eq2}) we evaluate the $\psi_D^{(D)}$ as a function of the quasi-power $\chi$. 
From there, we evaluate the injected power $I_D\equiv |\psi_D^{(D)}|^2$ which is further compared to the experimental value $P_\textrm{VNA}$. 
This comparison allow us to identify the appropriate quasi-power $\chi$ associated with the experimental incident power.
The other components of the defect mode $\psi_m^{(D)}$ can be also evaluated from Eq.~(\ref{Gfunction}) using the same steps as above. 
We have:
\begin{equation}\label{eq3}
\frac{\left(\psi_m^{(D)}\right)^2}{\chi}
=\left[-\frac{\bra{m}\G_0\ket{D}\bra{D}\G_0\ket{m}}{\diff{\bra{D}\G_0\ket{D}}{z}
}\right]_{z=\omega_D}.
\end{equation}
Knowledge of the $\omega_D$ and of the corresponding field amplitudes $\{\psi_m^{(D)}\}$ allow us to construct any other physical observable. 
The simplest one is the field intensity $I_D\equiv |\psi_D|^2$ at site $m_D=0$ which is associated with the pumped power and it is the controlled variable in the pump-probe experiment of Fig.~\ref{fig:fig1_photo_set_up}. 
In the specific case of the SSH with a lossy nonlinear defect, Eqs.~(\ref{eq:eq1},~\ref{eq:eq2}), and (\ref{eq3}) have been solved numerically by calculating the unperturbed Green's function $\G_0$ with  system size $N=17$. 
The theoretical results have been then compared with the experimental findings. 
Let us finally mention that in \cite{supplement} we provide an alternative derivation for the defect energy and the field amplitude of the corresponding mode which is based on an ansatz for the form of the defect mode. 

\begin{figure}
  \centering
  \includegraphics[width=1\columnwidth]{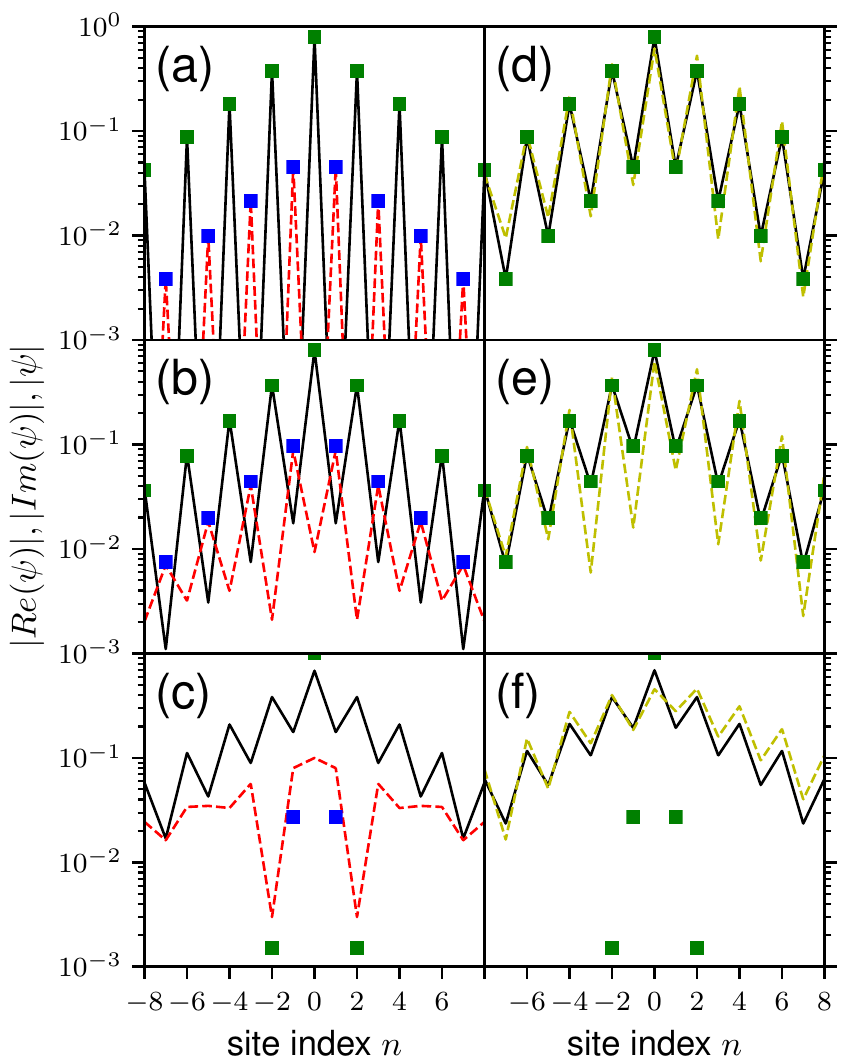}
  \caption{\label{fig:fig3_wv_green}
(Left column) 
Theoretical calculations for the ${\cal R}e\left(\psi_m^{(D)}\right)/{\cal I}m\left(\psi_m^{(D)}\right)$ using the exact (black line/red dashed line) and the approximated (green/blue squares) expression for $\Omega(I_D)$ from Eq.~(\ref{eq:Omega_sat}) for three pumping powers $P_\textrm{VNA}$ (a) -40\,dBm, (b) -10\,dBm, (c) 10\,dBm. 
(Right column) 
Experimental modulus profiles $\left|\psi_m^{(D)}\right|$ (yellow dashed lines) for three pumping powers $P_\textrm{VNA}$ (d) -40\,dBm, (e) -10\,dBm, (f) 10\,dBm. 
The theoretical calculations using an exact/approximate form for the nonlinearity are also reported with black solid lines/green squares. 
In all cases a normalization convention for wave functions ($\sum_m\left|\psi_m\right|^2=1$) is used for the presentation of the data.
}
\end{figure}

\emph{Results -- } 
A careful analysis of the experimental transmission spectrum of the SSH/Diode array and a comparison with the theoretical analysis of the spectrum allowed us to extract the nonlinear term $\Omega(I_D)$ appearing in Eq.~(\ref{eq:eq_nonl}). 
The best fit occurs for the expression:
\begin{equation}\label{eq:Omega_sat}
\Omega(I_D)=z_0-\frac{z_1}{1+ \alpha I_D}\approx (\beta_0 + \beta_1 I_D)
\end{equation}
where $z_0=(-40+18\imath)$\,MHz, $z_1=(-40+8\imath)$\,MHz, $\alpha=(1-2.8\imath)$\,mW$^{-1}$ have been extracted via comparison between the measured values of the defect frequency (see below) and the theoretical predictions (we have assumed that $\alpha_{\rm lin} =1$ i.e. $I_D=P_\textrm{VNA}$). 
The first equality is applicable for all measured powers that have been used in our experiments, while the second equality in Eq.~(\ref{eq:Omega_sat}) applies for moderate values of the pumped power up to $I_D\approx -10$\,dBm. 
In this pumped power regime, one can safely assume that $\beta_0\equiv \left(z_0 - z_1\right)=10i$\,MHz and $\beta_1\equiv z_1 \alpha$ has a negligible real part ({\cal R}e$[z_1 \alpha] \approx -17.6\,{\rm MHz/mW}\ll \epsilon/I_D$) and therefore can be considered (for any realistic purposes) to be imaginary.
Therefore, the system is expected to respect the \CT-symmetry (see previous discussion). 

We first present the parametric evolution of the defect frequency $\omega_D$ versus the pumped power $P_\textrm{VNA}$, see Fig.~\ref{fig:fig2_omega_shift}. 
The experimental values for $\omega_D$ (see filled cicles) have been extracted from the measured transmission spectra. 
A resonance peak in the middle of the band gap signifies the presence of the defect mode, and we have recorded the trace of such peak with increasing pumped power. 
At the same Fig.~\ref{fig:fig2_omega_shift} we also report the theoretical results for the defect frequency, extracted from the analysis of Eqs.~(\ref{eq:eq1},~\ref{eq:eq2}) using the nonlinear form Eq.~(\ref{eq:Omega_sat}). 
We find that for $P_\textrm{VNA} \leq -10$\,dBm the approximate form of $\Omega(I_D)$ in Eq.~(\ref{eq:Omega_sat}) captures all features of the experimental data. 
Consequently, the defect frequency $\nu_D={\cal R}e(\omega_D)$ is pinned to $\nu_0$, while its imaginary part $\gamma_D={\cal I}m(\omega_D)$ significantly grows with the pumped power $I_D$ due to the corresponding increase in the amount of nonlinear loss. 
Hence, the \CT-symmetry enforce a spectral (topological) protection against self-induced variations of the resonant frequency of the defect resonator. 
For even higher values of the pumped power $P_\textrm{VNA}$ (highlighted area in Fig.~\ref{fig:fig2_omega_shift}) the approximate expression for the nonlinearity Eq.~(\ref{eq:Omega_sat}) does not apply. 
Instead, a real (nonlinear) frequency shift of the defect resonator is self-induced which results to an explicit (self-induced) violation of the \CT-symmetry. 
Consequently, when $P_\textrm{VNA} \geq -10$\,dBm the defect frequency $\nu_D$ is not protected but rather shifts towards the band.

Next, we investigate the field profiles of the defect mode for various pumped powers $P_\textrm{VNA}$. 
In Fig.~\ref{fig:fig3_wv_green} (left column) we report the theoretical calculation of the field amplitudes $\psi_m^{(D)}$ for three representative values of $P_\textrm{VNA}=-40$\,dBm, $-10$\,dBm  and $10$\,dBm. 
We find that for pumped powers $P_\textrm{VNA} \leq -10$\,dBm, the defect mode respects the \CT-symmetry as it is evident from the fact that all even (odd)-sites are occupied by purely real (imaginary) wave function components. 
This analysis reconfirms our previous conclusion, which was based on the $\nu_D$ vs.~$P_\textrm{VNA}$ analysis: as long as the nonlinearity Eq.~(\ref{eq:Omega_sat}) is approximately purely imaginary, the defect mode is in the so-called {\it exact} \CT-symmetric phase i.e.\ it is also an eigenmode of the \CT-operator. 
In contrast, for higher pumped powers (see Fig.~\ref{fig:fig3_wv_green}c) the $\Omega(I_D)$ develops a considerable real part and the system experiences a self-induced \CT-symmetry violation. 
This is reflected in the fact that the real and imaginary parts of $\psi_m^{(D)}$ do not have any more a staggered form i.e. $\psi_m^{(D)}$ is not an eigenstate of \CT-operator ({\it broken phase}). 

The validity of the theoretical analysis of the field profile has been accessed via a direct comparison with the experimental measurements of the defect modulus profile $|\psi_m^{(D)}|$, see Fig.~\ref{fig:fig3_wv_green} (right column). 
The resulting agreement between the approximate form of $\Omega(I_D)$ and the ``actual" form is impressive for pumped powers $P_\textrm{VNA} \leq -10$\,dBms. 
This indicates that up to this pumped power the system (effectively) respects the \CT symmetry [see fig.~\ref{fig:fig3_wv_green}(a,b)] and is in good agreement with the experiment [see fig.~\ref{fig:fig3_wv_green}(d,e)]. 
Above $-10$\,dBm, the two forms of the nonlinearity provide different results indicating that the system has entered the self-induced explicit \CT-symmetric violation regime. 
Nevertheless, our theoretical calculations using the exact form of $\Omega(I_D)$ still agree nicely with the experimentally extracted wave profile [see fig.~\ref{fig:fig3_wv_green}(c,f)].

\emph{Conclusions -- } 
We have analyzed and demonstrated the topological properties of a nonlinear \CT-symmetric defect mode both theoretically and experimentally using a microwave platform that realizes a SSH CRMW array with a defect resonator coupled inductively to a PIN Diode. 
When the diode-induced nonlinearity is purely imaginary, the nonlinear defect mode is spectrally protected by the non-Hermitian \CT-symmetry. 
In particular, the defect frequency is in the middle of the band-gap while the field amplitude of the defect mode has a characteristic shape involving staggered imaginary and real parts. For high pumped powers, the nonlinearity acquires a sizable real part and the system experiences a self-induced explicit symmetry violation. 
In this case the defect mode is not any more protected by the \CT-symmetry. 
The self-induced \CT-symmetry violation can be an extremely desirable feature for various technological applications of topological photonics varying from topological protection of unidirectional defect modes at low incident powers to photonic reflective limiters.

\emph{Acknowledgments -- } 
T.K. and D.H.J. acknowledge partial support from ONR N00014-16-1-2803, from AFOSR via MURI grant FA9550-14-1-0037 and from NSF EFMA-1641109.


%%%%%%%%%%%%%%%%%%%%%%%%%%%%% Supplemental Material %%%%%%%%%%%%%%%%%%%%%%%%%%%%%%%%%
\onecolumngrid
\appendix
\newpage

\centerline{{\Large \bf Supplemental Material}}
\section{Non-Hermitian Green's Function}
\renewcommand\thefigure{S\arabic{figure}}    
\setcounter{figure}{0}
\setcounter{page}{1}
\renewcommand{\thepage}{S\arabic{page}}
\setcounter{page}{1}
\setcounter{equation}{0}
\renewcommand\theequation{S\arabic{equation}}

Let $\Hcal$ denote the Hamiltonian of a given system, which is assumed to be non-Hermitian. Also, let $\ket{\Psi_l}$ and $\bra{ \Phi_l}$ denote the $l$-th right and left eigenvectors of $\Hcal$, respectively i.e.
\begin{equation}
\Hcal\ket{\Psi_l}=\lambda_l\ket{\Psi_l} \qquad \textrm{and} \qquad \Hcal^\dagger\ket{ \Phi_l}=\lambda_l^*\ket{ \Phi_l},
\end{equation}
where $\lambda_l$ denotes the $l$-th eigenvalue of $\Hcal$. 
In fact, in case of non-degenerate Hamiltonians which are transpose-symmetric $\Hcal^T=\Hcal$ one can further show that
\begin{eqnarray}\label{eq:eq_transpose}
\Hcal^T \ket{\Phi_l^*}= \lambda_l \ket{\Phi_l^*};\quad
\Hcal \ket{\Phi_l^*}= \lambda_l \ket{\Phi_l^*};\quad
\ket{\Psi_l}&=& \ket{\Phi_l^*}
\end{eqnarray}

Analogous to orthonormality between eigenstates in Hermitian systems, we have biorthonormality for non-Hermitian systems, given by
\begin{equation}
\sum_l \frac{\ket{\Psi_l}\bra{ \Phi_l}}{\langle \Phi_l|\Psi_l\rangle}=1;\quad  \langle \Phi_l|\Psi_{l'}\rangle=\chi\delta_{l,l'} 
\end{equation}
We define the Green's function $\G$ of $\Hcal$ as
\begin{equation}
\G(z)\equiv \frac{1}{z-\Hcal},
\end{equation}
where $z$ is a complex variable. 
Using biorthonormality, we can re-express $\G$ as 
\begin{eqnarray}\nonumber
\G(z)=\frac{1}{z-\Hcal}
&=&\frac{1}{z-\Hcal}\sum_l \frac{\ket{\Psi_l}\bra{ \Phi_l}}{\langle \Phi_l|\Psi_l\rangle}\\
&=&\sum_l \frac{1}{\chi}\frac{\ket{\Psi_l}\bra{ \Phi_l}}{z-\lambda_l}.
\end{eqnarray}
As a result, we obtain the following expression for the matrix elements of $\G$ in Wannier basis.
\begin{eqnarray}\nonumber
\bra{m}\G(z)\ket{m}
&=&\sum_l\frac{1}{\chi} \frac{\braket{m}{\Psi_l}\braket{ \Phi_l}{m}}{z-\lambda_l}\\
&=&\sum_l \frac{1}{\chi}\frac{\psi_m^{(l)}  (\phi_m^{(l)})^*}{z-\lambda_l}=\sum_l \frac{1}{\chi}\frac{(\psi_m^{(l)})^2}{z-\lambda_l}
\end{eqnarray}
where $\varphi_m^{(l)}$ and $ \chi_m^{(l)}$ denote the wave function of each corresponding eigenstate at the $m$-th site. 
At the last equality we have used that $\Hcal^T=\Hcal$ and therefore Eq.~(\ref{eq:eq_transpose}) holds. 
At this point, we can immediately see that the Green's function $\G$ has simple poles at "discrete" eigenvalues $\lambda_l$ of $\Hcal$, where discrete eigenvalues imply the presence of bound states. Furthermore, the residue of the $m$-th diagonal element of $\G$ at $z=\lambda_l$ is equal to the squared wave function of the bound state $\ket{\Psi^{(b)}}$ at the $m$-th position.

\section{Ansatz Method for Evaluating the Defect Mode}

In the main text we have used a non-Hermitian Green's function approach in order to investigate the properties of the nonlinear defect mode. 
We could, however, utilize an ansatz approach, which has the benefit of being more transparent, and leads to an analytical (transcendental) solution for the defect mode. 
To this end, we consider a general nonlinearity $\Omega(I_D)$ which applies only on the defect resonator. 
Moreover, we assume that the number of sites $N\to\infty$ (thermodynamic limit).

We proceed using the following ansatz for the defect mode:
\begin{equation}
\psi^{(D)}_m=\left\{ 
\begin{array}{ll}
A\e^{\imath \mu \frac{|m|}{2}} & \textrm{for even } m\\
B\e^{\imath \mu \frac{|m|-1}{2}} & \textrm{for odd } m\\
\end{array}
\label{ansatz}
\right.\,,
\end{equation}
where $\mu$ is in general complex and will determine the decay rate of the defect mode away from the nonlinear defect site.
Substituting Eq.~(\ref{ansatz}) in Eq.~(\ref{eq:CMTeqs}), for $m = m_D = 0$, $m = 1$, and $m = 2$ we get the following set of coupled equations
\begin{equation}\label{eq:CMEs}
\left\{ 
\begin{array}{lcl}
(\omega - \Omega(I_D)) A &=& 2t_2 B\\
\omega B &=& t_2 A + t_1A\e^{\imath \mu}\\
\omega A \e^{\imath \mu}&=& t_1 B + t_2B\e^{\imath \mu}\,.
\end{array}
\right.
\end{equation}
From the first of Eqs.~(\ref{eq:CMEs}), we obtain $B = \frac{(\omega - \Omega(I_D))A}{2 t_2}$. 
Using this relation, in the second Eq.~(\ref{eq:CMEs}) and after trivial algebraic manipulations we get
\begin{equation}\label{eq:ansatz_eq1}
\omega^2 - \omega \Omega(I_D)
= 2 {t_2}^2 + 2 t_1 t_2 \e^{\imath \mu}
\end{equation}
In the same manner, using the expression for $B$, in the third line of Eq.~(\ref{eq:CMEs}) we get 
\begin{equation}\label{eq:ansatz_eq2}
2 t_2 \omega \e^{\imath \mu} = (t_1 + t_2 \e^{\imath \mu}) (\omega - \Omega(I_D))\,.
\end{equation}
By introducing the variable $x = \e^{\imath \mu}$ and noting that $I_D = |A|^2$, Eq.~(\ref{eq:ansatz_eq1}) and Eq.~(\ref{eq:ansatz_eq2}) can be solved for $\omega$ and $x$ for a fixed value of $I_D$. 
Note that the evaluation of $\omega$ and $x$ determines the wave function $\psi^{(D)}_m$. 
The only undetermined factor is the global complex phase of the wave function $\psi^{(D)}_m$, which does not have a physical meaning and thus may be arbitrarily chosen.

\begin{figure}
	\centering
	\includegraphics[width=.99\linewidth]{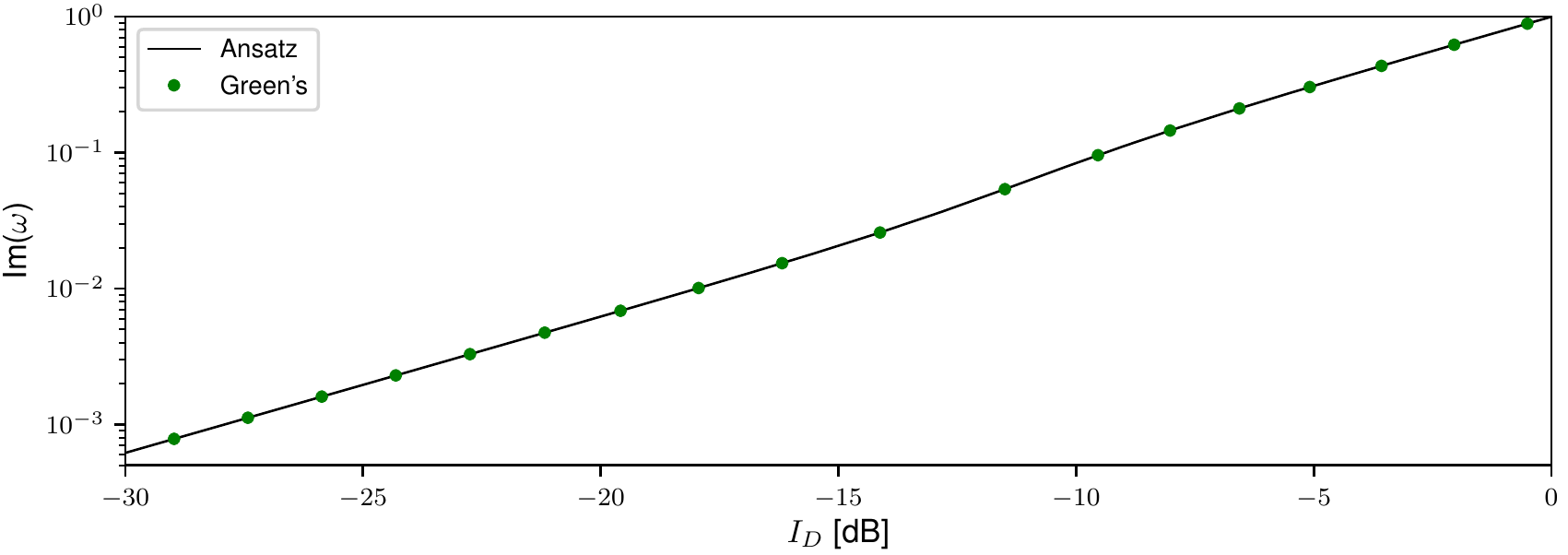}
	\caption{\label{fig:supp1}
		Defect mode frequency $\omega$ (measured in GHz), purely imaginary, with varying local field intensity $I_D$ (expressed as dBm) at the defect site. 
		Results calculated using the two independent methods are compared, where the black line is obtained from the ansatz-based approach, and the green dots from the Green's function approach.
	}
\end{figure}

\begin{figure}
	\centering
	\includegraphics[width=.99\linewidth]{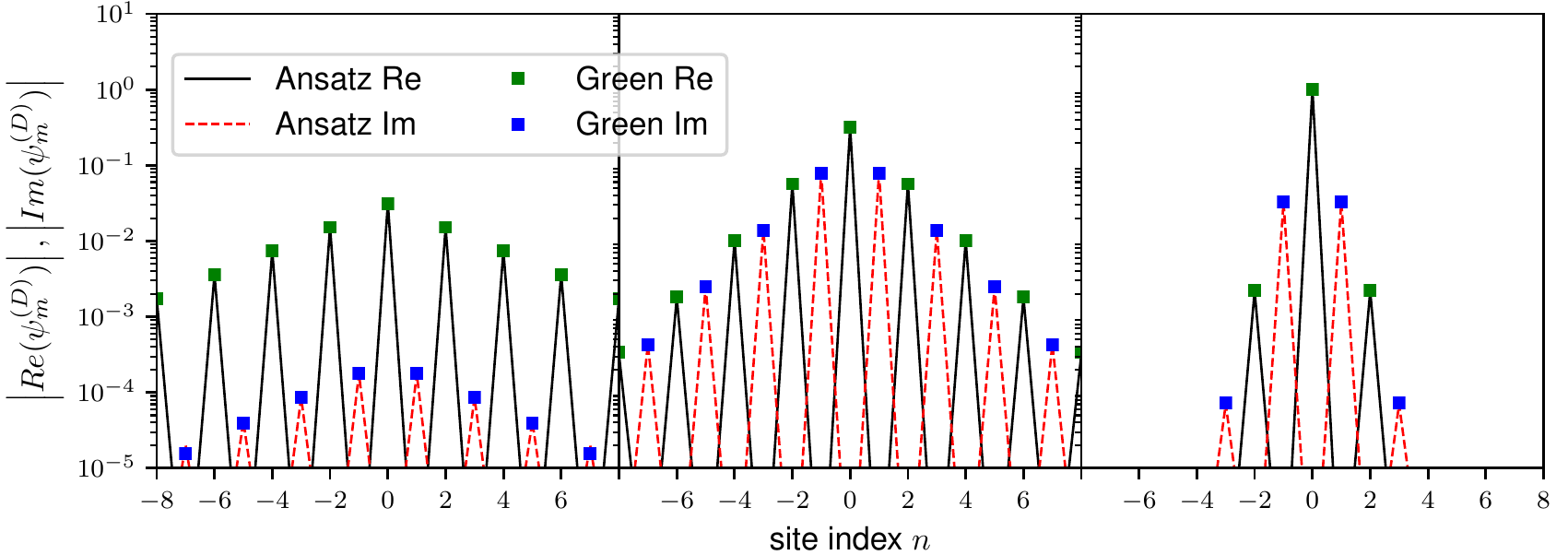}
	\caption{\label{fig:supp2}
		Complex wave profile $\psi^{(D)}_m$ of the defect mode for varying local field intensity $I_D$ (increasing from left to right; $-30$\,dBm, $-10$\,dBm, $0$\,dBm).
		Results have been calculated using the ansatz-based approach (black lines for the real part; red lines for the imaginary part) and the Green's function approach (green squares for the real part; blue squares for the imaginary part).
		Only the magnitudes of the real and imaginary parts are presented.
	}
\end{figure}

For comparison with the Green's function approach, we consider the particular case of a two photon absorption (TPA) nonlinearity
\begin{equation}
\Omega(I_D) = \imath I_D
\end{equation}
which is a good approximation for $P_\textrm{VNA} \leq -10$\,dBm (see main text).
Having the freedom to choose the global complex phase of $\psi^{(D)}_m$, we assume that $A\in\mathbb{R}$ and $A > 0$.
Since $\Omega(I_D)$ is purely imaginary in this case, the Hamiltonian of the system preserves $\CT$ symmetry for any $I_D$.
Therefore, we expect that the chiral symmetric defect mode in the low $I_D$ limit will become the $\CT$ symmetric defect mode for any $I_D$.

The results derived from the two independent methods show a perfect match (See Fig.~\ref{fig:supp1} and Fig.~\ref{fig:supp2}), despite the fact that the Green's function has been numerically calculated for a finite system size $N = 17$. 
Moreover, both methods show that the nonlinear defect mode nicely captures all conditions required for an exact $\CT$-symmetric mode.

\begin{thebibliography}{39}
\bibitem [{not()}]{note1}%
	\BibitemOpen
	\href@noop {} {}\bibinfo {note} {In the special case where $\Omega(I_D)$ is
		negligible, the Hamiltonian respects chiral symmetry, i.e. the Hamiltonian
		$\Hcal$ anti-commutes with the $\C$-operator. Chiral symmetric states
		($\C\ket{\Psi}=\ket{\Psi}$) have fixed eigenvalues at zero ($\omega=0$), and
		their mode profiles show a characteristic staggered shape ($\psi_m = 0$ for
		all odd $m$).}\BibitemShut {Stop}%
\end{thebibliography}
\end{document}